\documentclass[aps,prl,twocolumn,superscriptaddress,showpacs]{revtex4}
\usepackage{graphicx}
\newcommand{\simlt}  {\raisebox{-.6ex}{$\stackrel{\textstyle <}{\sim}$}}

\newcommand{\J} {J_{CP}}
\newcommand{\nua} {\nu_1}
\newcommand{\nub} {\nu_2}
\newcommand{\nuc} {\nu_3}
\newcommand{\nue} {\nu_e}
\newcommand{\num} {\nu_{\mu}}
\newcommand{\nut} {\nu_{\tau}}
\newcommand{\Uea} {U_{e1}}
\newcommand{\Ueb} {U_{e2}}
\newcommand{\Uec} {U_{e3}}
\newcommand{\Uma} {U_{\mu 1}}
\newcommand{\Umb} {U_{\mu 2}}
\newcommand{\Umc} {U_{\mu 3}}
\newcommand{\Uta} {U_{\tau 1}}
\newcommand{\Utb} {U_{\tau 2}}
\newcommand{\Utc} {U_{\tau 3}}
\newcommand{\Umns} {U}
\newcommand{\reuec} {{\rm Re}(\Uec)}
\newcommand{\imuec} {{\rm Im}(\Uec)}
\newcommand{\mat}[9] {\left( \matrix{#1 & #2 & #3 \cr
                                     #4 & #5 & #6 \cr 
                                     #7 & #8 & #9 \cr} \right)}

\newcommand{\mt}{$\mu$-$\tau$}

\newcommand{\R}{C}
\newcommand{\SASQ}{\sin^2\!{\Delta_{31}}}

\newcommand{\SA}{\sin\!{\Delta_{31}}}

\newcommand{\nnm}{\nonumber}

\begin{document}

%\preprint{}

\title{Simplified Unitarity Triangles for the Lepton Sector}



\author{James~D.~Bjorken}
\email[]{bjorken@slac.stanford.edu}
\affiliation{Stanford Linear Accelerator Center, Stanford University, 
Stanford, California 94309, USA}
\author{P.~F.~Harrison}
\email[]{p.f.harrison@warwick.ac.uk}
\affiliation{Department of Physics, University of Warwick, Coventry CV4 
7AL, UK}
\author{W.~G.~Scott}
\email[]{w.g.scott@rl.ac.uk}
\affiliation{CCLRC Rutherford Appleton Laboratory, Chilton, Didcot, 
Oxon OX11 0QX, UK}

\date{\today}

\begin{abstract}
Encouraged by the latest SNO results, we consider the lepton mixing 
matrix in the approximation that the $\nub$ mass eigenstate is 
trimaximally (democratically) mixed. This suggests a new parameterization 
of the remaining mixing degrees of freedom, which eschews mixing angles, 
dealing instead, directly with the complex parameter $\Uec$ of the mixing 
matrix. Unitarity triangles then take a particularly simple form,
which we hope will faciltate comparison with experiment.
\end{abstract}

%\keywords{}

\maketitle

Recent years have seen huge advances in our knowledge of the 
properties of neutrinos. Most recently, SNO \cite{SNO:1,SNO:2} has provided the 
best evidence for neutrino flavour change, which, coupled with evidence 
from atmospheric neutrinos \cite{ATMOS:1,ATMOS:6}, reactors 
\cite{REACTOR:1,REACTOR:4,KAMLAND} and accelerator experiments 
\cite{K2K}, has enabled the basic form of the MNS \cite{MNS:1} 
lepton mixing matrix, $\Umns$, to be determined \cite{FITS:1}.
%we review below.

Atmospheric neutrino data \cite{ATMOS:1,ATMOS:6}, together with K2K 
\cite{K2K} and reactor data \cite{REACTOR:1,REACTOR:4}, give (at 68\% CL)\cite{FITS:1}:
\begin{equation}
|\Uec|^2\,\simlt\,0.013 \qquad |\Umc|^2=0.50\pm 0.11.
\label{um3}
\end{equation}
Therefore $|\Utc|^2 \approx 0.50\pm 0.11$, implying that
\begin{equation}
|\Uec|^2 \approx 0 \qquad |\Umc|\approx |\Utc| \approx \frac{1}{\sqrt{2}}.
\end{equation}
We may choose the phases of $\num$ and $\nut$ such that to a good 
approximation
\begin{equation}
\nuc = \frac{1}{\sqrt{2}}(\num - \nut).
\label{tbmnu3}
\end{equation}
Unitarity then implies that, with the above choice of phases,
\begin{equation}
\Umb \approx \Utb \qquad \Uma \approx \Uta,
\label{um2ut2}
\end{equation}
ie.~approximate \mt\ symmetry \cite{MUTAUSYMM:7,MUTAUSYMM:1,MUTAUSYMM:6}.
From the analysis of solar neutrino data \cite{SOLAR:1,SOLAR:2,SOLAR:3,SOLAR:4,SOLAR:5}, especially from 
SNO \cite{SNO:1,SNO:2}, we have (at 68\% CL):
\begin{equation}
|\Ueb|^2 = 0.31 \pm 0.04
\label{ue2}
\end{equation}
implying
\begin{equation}
|\Ueb| \approx |\Umb| \approx |\Utb| \approx \frac{1}{\sqrt{3}}.
\label{nu2approx}
\end{equation}
The phases of $\nub$ and $\nue$ can be chosen such that
\begin{equation}
\nub =  \frac{1}{\sqrt{3}}(\nue + \num + \nut).
\label{tbmnu2}
\end{equation}
Unitarity now fixes the remaining three MNS matrix 
elements up to their overall phase, which may be chosen so that
\begin{equation}
\nua =  \frac{1}{\sqrt{6}}(2\nue -\nut - \num).
\label{tbmnu1}
\end{equation}
Both the relative precision, and the absolute precision, in 
the determination of $|\Ueb|^2$ in Eq.~(\ref{ue2}) is better than that of 
$|\Umc|^2$ in Eq.~(\ref{um3}), making $|\Ueb|\approx \frac{1}{\sqrt{3}}$ 
currently the best-determined of the MNS matrix elements.

Equations (\ref{tbmnu3}), (\ref{tbmnu2}) and (\ref{tbmnu1}) together 
define the tribimaximal mixing texture \cite{TRIMAX:3,TBM:1,VENICE}, which 
we may take to be the leading approximation to the lepton mixing matrix. 
This texture is clearly evocative of symmetries at work. 
Taking the neutrino flavour 
eigenstates to define the orientation of a cube, the $\nu_2$ eigenstate, 
Eq.~(\ref{tbmnu2}), lies along the body diagonal of the cube, while
the $\nuc$ mass eigenstate lies in the plane of the $\num-\nut$ face, at 
$45^{\circ}$ to the $\num$ state.
It may be remarked that the same mixing matrix elements also occur 
as the $M=0$ subset of the $j\times j =1 \times 1$ set of Clebsch-Gordan 
coefficients.

An extensive future experimental neutrino program is being planned 
\cite{NEWEXPT:1,NEWEXPT:2,NEWEXPT:3} which will 
refine the tribimaximal picture outlined above. 
The current situation for neutrino physics and the MNS matrix appears 
analogous to that earlier for $B$ physics and the CKM matrix, in which 
the leading approximation to the matrix was established experimentally, 
long before its smallest elements were determined. 
In that case, the Wolfenstein 
parametrization has become widely adopted \cite{WOLFENSTEIN}. This 
approximate parameterization avoids the introduction 
of mixing angles, dealing instead directly with the elements of the 
mixing matrix. Visualisation of its complex elements 
has been facilitated with the help of ``unitarity triangles'' \cite{UT:1,UT:2,UT:3,UT:4,UT:5,UT:6,UT:7}, which, in their normalised form
\cite{PDG} use only two parameters.
It is our purpose here to propose a simple two-parameter approximation  
for the MNS mixing in the lepton sector. It is motivated by the 
phenomenological success of tribimaximal mixing, which it takes as a 
starting point, but does not in fact depend upon this mixing form being 
exact, or even theoretically consequential.

In the CKM case, phases are chosen such that the 
diagonal elements, as well as the two elements in the row above the 
diagonal, are chosen real and positive. Then unitarity 
controls the remaining elements, with the free variable of greatest 
importance being $V_{ub}$. 
The values of the two elements in the row below 
the diagonal are insensitive to details, while the remaining one, 
$V_{td}$, is related to $V_{ub}$ via the well-known unitarity 
triangle.

For the case of the MNS matrix, the optimal choice of phases is different 
because the mixing pattern is different. The most striking feature of the 
data is arguably the near-democratic mixing of the $\nu_2$ mass eigenstate, 
as expressed in Eq.~(\ref{nu2approx}). 
As we show later, even for subdominant oscillation phenomena,
it is a robust approximation to assume exact trimaximal mixing for 
the $\nu_2$ column.
A natural choice of phases is then to require the $\nu_2$ column to be 
real and positive.
The remaining elements are then controlled by $U_{e3}$ (the analogue of 
$V_{ub}$), which is again small, and again vital to the unresolved issue of $CP$ violation, this time within the lepton sector.

We construct a generalisation \cite{SYMMSGENS,CHARACTERS:1,SIMPLEST} of the tribimaximal form, consistent with the
above considerations:
\begin{eqnarray}
\Umns&\simeq&
\left( \matrix{ 
  \frac{2}{\sqrt{6}} &  \frac{1}{\sqrt{3}}  &   0  \cr
 -\frac{1}{\sqrt{6}} &  \frac{1}{\sqrt{3}}  & \frac{1}{\sqrt{2}}   \cr
 -\frac{1}{\sqrt{6}} &  \frac{1}{\sqrt{3}}  & -\frac{1}{\sqrt{2}}    \cr
 }\right)
\mat
  {\R}{0}{\sqrt{\frac{3}{2}}U_{e3}} 
  {0}{1}{0}
  {-\sqrt{\frac{3}{2}}U^*_{e3}}{0}{\R}\nonumber\\
%\label{expan1}\\
&=&\mat
{\frac{2}{\sqrt{6}}\R}
{\frac{1}{\sqrt{3}}}
{U_{e3}}
{-\frac{1}{\sqrt{6}}\R-\frac{\sqrt{3}}{2}U^*_{e3}}
{\frac{1}{\sqrt{3}}}
{\frac{1}{\sqrt{2}}\R-\frac{{1}}{2}U_{e3}}
{-\frac{1}{\sqrt{6}}\R+\frac{\sqrt{3}}{2}U^*_{e3}}
{\frac{1}{\sqrt{3}}}
{-\frac{1}{\sqrt{2}}\R-\frac{{1}}{2}U_{e3}}
\label{exactparam}
\end{eqnarray}
where 
\begin{equation}
\R=\sqrt{1-\frac{3}{2}|U_{e3}|^2}\simeq 1.
\label{cdef}
\end{equation}
The matrix in Eq.~(\ref{exactparam}) is exactly unitary, by construction. 
The need for the 
factor $\sqrt{\frac{3}{2}}$ in the second matrix in the expansion of Eq.~(\ref{exactparam}) is to 
keep $U_{e3}$ itself as the operative parameter in the resulting product. 
Dropping terms of order $|U_{e3}|^2$ for simplicity (ie. setting $\R=1$), 
we obtain the approximation
\begin{eqnarray}
U&\simeq&
\mat{\frac{2}{\sqrt{6}}}{\frac{1}{\sqrt{3}}}{U_{e3}}
{-\frac{1}{\sqrt{6}}-\frac{\sqrt{3}}{2}U^*_{e3}}{\frac{1}{\sqrt{3}}}
{\frac{1}{\sqrt{2}}-\frac{{1}}{2}U_{e3}}
{-\frac{1}{\sqrt{6}}+\frac{\sqrt{3}}{2}U^*_{e3}}{\frac{1}{\sqrt{3}}}
{-\frac{1}{\sqrt{2}}-\frac{{1}}{2}U_{e3}}.
\label{newparam}
\end{eqnarray}

In our parameterization, $|U_{e3}|$ corresponds exactly to the same quantity 
in the standard (PDG) parameterization \cite{PDG}:
\begin{eqnarray}
|U_{e3}|=\sin{\theta_{13}},
\label{ue3s13}
\end{eqnarray}
while our phase convention for $\Uec$ differs only slightly from the usual 
one. This phase is fixed by the construction in Eq.~(\ref{exactparam}) 
such that $\delta'\equiv -{\rm Arg}(U_{e3})$ is given by
\begin{eqnarray}
\sin{\delta'}&=&\sin{2\theta_{23}}\sin{\delta}\label{phase}\\
&\simeq&\sin{\delta}
\label{phases}
\end{eqnarray}
(this explains our choice of sign for the $\nuc$ column). Exact 
equivalence of these two phase definitions is obtained in the two special 
cases: ${\rm Re}(U_{e3})=0$ or ${\rm Im}(U_{e3})=0$.

$CP$ violation is, of course, governed by ${\rm Im}(U_{e3})$. We have for 
Jarlskog's $CP$-violating observable \cite{JCP:1,JCP:2}:
\begin{equation}
\J=\frac{\R\,{\rm Im}(U_{e3})}{3\sqrt{2}}\simeq \frac{{\rm Im}(U_{e3})}{3\sqrt{2}}.
\label{eqjcp}
\end{equation}
The parameter ${\rm Re}(U_{e3})\simeq \sqrt{2}\,(\frac{\pi}{4}-\theta_{23})$ clearly violates \mt\ reflection symmetry 
(simultaneous interchange of $\mu$ and $\tau$ flavour labels and a $CP$ transformation \cite{MUTAUSYMM:1}).

\begin{figure}
\includegraphics[width=9cm]{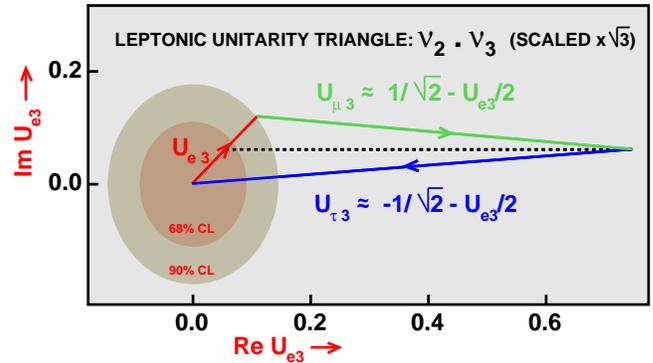}
\caption{\label{trianglesC} The ``$\nub.\nuc$'' unitarity triangle 
representing the orthogonality of the $\nub$ and $\nuc$ mass eigenstates 
(the dashed line bisects the $U_{e3}$ side, is real and has length 
$\frac{1}{\sqrt{2}}\R$). The triangle has been scaled by a factor $\sqrt{3}$ 
so that the sides are equal to the components of the $\nuc$ column of the 
lepton mixing matrix. The error-ellipse indicates the allowed range of 
$|\Uec|$: the vertical error is given by the reactor bound 
\cite{REACTOR:1,REACTOR:4}, while the horizontal error was obtained by 
combining this with the implied constraint from the experimental value of 
$|\Umc|$ (Eq.~(\ref{um3})), under the assumption of exact trimaximal mixing 
of the $\nub$ mass eigenstate.}
\end{figure}

In the $CP$-violating case, leptonic unitarity triangles \cite{UT:8,UT:9,UT:10} may be constructed 
using the orthogonality of pairs of columns (or rows)
of the mixing matrix. With our parameterization, 
Eqs.~(\ref{exactparam})-(\ref{newparam}),
most of the triangles become 
especially simple, eg.~the ``$\nub.\nuc$'' 
triangle shown in Fig.~\ref{trianglesC}, expressing the 
orthogonality of the $\nub$ and $\nuc$ columns. Its sides are all 
proportional to the elements of the $\nuc$ 
state (with a common factor $1/\sqrt{3}$ removed 
by rescaling).
%
%This is a result of the trimaximally 
%mixed $\nub$ column of our mixing matrix, and our choice of phases.
This is perhaps the most useful of the six possible triangles, as one of 
its sides is simply proportional to $U_{e3}$ itself, and is hence the 
leptonic analogue of the most commonly used ``$d.b$'' 
triangle of the quark sector.

The triangle expressing the orthogonality of 
the $\nua$ and $\nub$ columns has a similar simplification to that 
of the $\nub.\nuc$ triangle discussed above, 
its sides all being proportional to the elements 
of the $\nua$ mass eigenstate. It may be constructed from a (real) line 
of length $\Uea=\frac{2}{\sqrt{6}}C$, bisected by a (complex) line given by 
$\frac{\sqrt{3}}{2}\Uec$ whose end defines the third vertex.
The triangle expressing the orthogonality of 
the $\nua$ and $\nuc$ columns is the least degenerate of the three 
column-based triangles, but has no special simplification, in general, here.

The three triangles representing orthogonality of the rows of the 
MNS matrix are now similarly determined entirely in terms of 
$U_{e3}$. They each share one angle with each of the column-wise 
triangles. Note that with $\nu_2$ exactly trimaximally mixed
each of these row-based triangles stands
on a common real base of length $\frac{1}{3}$,
which could be rescaled to unity if desired.

Of course, the rephasing invariance of observables means that only 
the triangles' opening angles, side-lengths, and areas have physical 
significance (see below), their orientation being irrelevant.
Symmetries are reflected in the shapes of the triangles, 
eg.~with exact \mt\ reflection symmetry, all three column-based triangles 
are isosceles 
(their sides carry lepton flavour indices), while the $\nue.\nut$ 
and $\nue.\num$ triangles 
become congruent to each other. If $CP$ is conserved, each 
triangle reduces to a line.

We turn now to the use of the triangles in the phenomenology of leptons, 
especially neutrino oscillations.
The sides of the row-based triangles are 
given by complex products of MNS matrix elements,
$U_{\alpha i}U_{\beta i}^*$, where
the flavour labels $\alpha$ and $\beta$ define the triangle, and the mass 
index $i$, labels the side.
These triangle sides are the (complex) magnitudes of the three 
sub-amplitudes which mediate neutrino oscillations between flavour 
$\alpha$ and flavour $\beta$, one for each neutrino mass eigenstate, $i$:
\begin{eqnarray}
|A(\nu_\alpha\to\nu_{\beta})|&=&|\sum_{i=1,3}\,U_{\alpha i}U_{\beta i}^*\,{\rm e}^{(-im_i^2L/2E)}\,|
\label{appearAmp}\\
&=&2\,|U_{\alpha i}U_{\beta i}^*\sin\Delta_{ik}
+U_{\alpha j}U_{\beta j}^*\sin\Delta_{jk}{\rm e}^{i\Delta_{ij}}|\nnm
\end{eqnarray}
($\alpha\neq\beta$, $i\neq j\neq k$), where 
$\Delta_{ij}\equiv(m^2_i-m^2_j)L/4E$, $L$ is the propagation 
distance, $E$ is the neutrino energy and $m_i$ is the mass of $\nu_i$.
Squaring Eq.~(\ref{appearAmp}) to obtain appearance probabilities, 
we see that the triangles' side-lengths, 
$|U_{\alpha i}U_{\beta i}^*|$, and their external angles, 
$\phi_{\gamma k}\equiv{\rm Arg}(U_{\alpha i}U^*_{\beta i}U^*_{\alpha j}U_{\beta j})$, 
($\gamma\neq \alpha$ etc.)\footnote
{The $\phi_{\alpha i}$ notation, serves equally for both sets of 
triangles: for row-based triangles, the index $\alpha$ 
defines the triangle representing the orthogonality
of eigenstates $\nu_{\beta}$ and $\nu_{\gamma}$, while the index $i$ 
labels the triangle side opposite the angle in question. For column-based 
triangles, the interpretations of the first and second indices are 
exchanged. The same matrix of 9 angles describes both sets of triangles.},
appear explicitly (see eg.~Eq.~(\ref{pmue}) below).
With our parameterization,
$U_{\alpha 2}U_{\beta 2}^*=1/3$ for all flavour-pairs, $\alpha$, $\beta$, 
while other side-lengths become simple functions of $\Uec$.

The $CP$ parameter, 
$\J=|U_{\alpha i}U^*_{\beta i}||U_{\alpha j}U^*_{\beta j}|\sin{\phi_{\gamma k}}$, 
$\forall\, \gamma, k$, equals twice the area of each (un-rescaled) 
triangle. For the triangle in Fig.~\ref{trianglesC}, we find 
using Eq.~(\ref{eqjcp}):
\begin{equation}
\sin{\phi_{\alpha 1}}=\frac{\R\,\imuec}{\sqrt{2}\,|U_{\beta 3}|\,
|U_{\gamma 3}|}\simeq\frac{\imuec}{\sqrt{2}\,|U_{\beta 3}|\,|U_{\gamma 3}|}
\label{angles3}
\end{equation}
with $\alpha\neq\beta\neq\gamma$. Expressions may similarly be obtained 
for the angles of all the other triangles.

Expressions for neutrino oscillation survival probabilities involve 
directly the moduli-squared of MNS matrix elements, 
$|U_{\alpha i}|^2$. 
Here again, $|U_{\alpha 2}|^2=1/3$ for all flavours $\alpha$.
For $i=3$, these are
the side lengths of the $\nub.\nuc$ triangle of Fig.~\ref{trianglesC} 
which (for non-trivial cases) are given by:
\begin{eqnarray}
|U_{\alpha 3}|^2
&=&\frac{1}{2}-\frac{|U_{e3}|^2}{2}\mp\frac{\R\,\reuec}{\sqrt{2}}\nnm\\
&\simeq&\frac{1}{2}\mp \frac{\reuec}{\sqrt{2}},\label{sides3}
\end{eqnarray}
where the upper(lower) signs correspond to $\alpha=\mu(\tau)$, 
respectively (the approximate form may be obtained directly 
by inspection of Fig.~\ref{trianglesC}).
The analogous result for $i=1$ is obtained from the 
$\nua.\nub$ triangle as:
\begin{eqnarray}
|U_{\alpha 1}|^2&=&\frac{1}{6}+\frac{|U_{e3}|^2}{2}\pm 
\frac{\R\,{\rm Re}(U_{e3})}{\sqrt{2}}\nnm\\
&\simeq&\frac{1}{6}\pm\frac{{\rm Re}(U_{e3})}{\sqrt{2}}.
\label{sides1}
\end{eqnarray}

A new generation of experiments is being planned and built 
to address the question of non-leading neutrino oscillations 
\cite{NEWEXPT:1,NEWEXPT:2,NEWEXPT:3}. 
The description of the observables at these new experiments 
is usefully presented as a series expansion 
\cite{AKHMEDOV} in the two small quantities:
\begin{eqnarray}
\alpha\equiv\frac{\Delta_{21}}{\Delta_{31}}\simeq 0.026,
\,\, 0.021\,\simlt\,\alpha\,\simlt\,0.036\,\,({\rm 90\%~CL}),
\label{smallpars}
\end{eqnarray}
(describing the hierarchy of mass-squared differences), and 
$|U_{e3}|=\sin\theta_{13}\,\simlt\,0.17~({\rm 90\%~CL})$, 
as a function of the standard parameters. In terms of our parameterization, we 
find a considerable simplification. To leading (second) order in small
quantities, the $\nu_e$ appearance probability in a long baseline muon
neutrino beam is (in vacuum):
%\newpage
%
\begin{widetext}
\begin{eqnarray}
P(\num\rightarrow\nue)\simeq2|\Uec|^2\SASQ+\frac{4}{9}\alpha^2\Delta_{31}^2
+\frac{8\alpha|\Uec|}{3\sqrt{2}}
\Delta_{31}\SA
\cos{(\Delta_{31}+\delta')}
\label{pmue}
\end{eqnarray}
\end{widetext}
where the $CP$-violating phase 
$\delta'\simeq\delta\simeq-\phi_{\tau 1}$
at this order of approximation (see Fig.~\ref{trianglesC} and 
Eqs.~(\ref{phase})-(\ref{phases})). 
The electron anti-neutrino disappearance probability at a (not too 
distant) reactor experiment is given, again to leading (second) order in small 
quantities, by:
\begin{eqnarray}
1-P(\overline{\nu}_e\rightarrow\overline{\nu}_e)
\simeq4|\Uec|^2\SASQ+\frac{8}{9}\alpha^2\Delta_{31}^2.
\label{pee}
\end{eqnarray}
Although Eqs.~(\ref{pmue}) and (\ref{pee}) have been derived under the 
assumption of exact trimaximal mixing of the $\nub$ mass eigenstate, 
it can be shown that they are still valid at second order in small 
quantities, even in the case that this assumption is broken by small 
perturbations. This justifies our proposal of Eq.~(\ref{newparam}) 
as a viable approximate parameterization of the MNS matrix.

As well as oscillation phenomenology, the row-based triangles play a role 
in (lepton number violating) radiative decays of charged leptons, while
the column-based ones play an analogous role in radiative decays of
neutrinos, although admittedly, these are unlikely to play a major role in 
phenomenology, as they are all highly suppressed in the 
Standard Model \cite{BOEHMANDVOGEL:1}.

Clearly, neutrino oscillation experiments have come a long way since one 
could countenance the idea of trimaximal mixing for all three neutrino 
species \cite{TRIMAX:1,TRIMAX:2,TRIMAX:3}. 
There is today however, still the real prospect that the $\nub$ mass 
eigenstate is indeed trimaximally mixed, and this 
hints at deeper symmetries beyond the Standard Model. While it is 
important to test experimentally whether the $\nub$ mass 
eigenstate is {\em exactly} trimaximally mixed, we have argued that this
assumption anyway provides a simplifying and illustrative approximation to 
the phenomenology of neutrino oscillations.

\begin{acknowledgments}
This work was supported by the UK Particle Physics and Astronomy Research Council 
(PPARC), and in part, by the US. Department of Energy contract DE-AC02-76SF00515. 
Two of us (PFH and JDB) acknowledge the hospitality of the Centre for 
Fundamental Physics (CfFP) at CCLRC Rutherford Appleton Laboratory.
\end{acknowledgments}
\bibliography{citations}
\end{document}